\documentclass[9pt,twocolumn,twoside]{osajnl}

\journal{ol} 

\setboolean{shortarticle}{true}

\title{All-fiber laser pump reflector based on a fs-written VBG}

\author[*]{Lauris Talbot}
\author[ ]{Pascal Paradis}
\author[ ]{Martin Bernier}

\affil[ ]{Centre d'optique, photonique et laser (COPL), Université Laval, Québec, Canada, G1V 0A6}

\affil[*]{Corresponding author: lauris.talbot.1@ulaval.ca}

\begin{abstract}
We report on the writing of chirped volume Bragg gratings using 400 nm femtosecond pulses and the phase-mask inscription technique in the pure silica inner cladding of the gain fiber. The usefulness of such fiber component is demonstrated as a pump reflector in a cladding-pumped all-fiber laser. A reflectivity of 53\% for the residual pump power guided in the highly multimode 125 \textmu m fiber’s cladding is achieved, allowing for significant pump recycling in the fiber laser cavity which therefore optimizes the overall laser performances. As proof of concept, adding this pump reflector at the output of a 25 W erbium-doped fiber laser operating near 1.6 \textmu m increased its slope efficiency according to the launched pump power from 17\% to 23\%, with the beneficial effect of reducing by 25\% the optimal gain fiber length. This new fiber laser component would have a great impact on the efficiency and cost of high-power cladding-pumped fiber laser systems.
\end{abstract}

\setboolean{displaycopyright}{true}

\begin{document}

\maketitle
The cladding-pumping scheme for rare-earth doped fiber lasers has made their power scaling possible thus opening their use for many high-end applications \cite{zervas2014high, richardson2010high, gapontsev20052}. This approach allows to easily couple low brightness and highly multimode, but very powerful pump light into the cladding of the active fiber that is absorbed along the fiber and converted to a core guided laser signal with an excellent beam quality and therefore a very large brightness. The reduction of the pump absorption rate distributes the gain along much longer lengths in comparison to core-pumping schemes which facilitates greatly the thermal dissipation of the heat generated during such process. However, laser cavity lengths have to be particularly long to achieve substantial pump absorption. The laser cavity becomes consequently more lossy and expensive, and starts to be sensitive to nonlinear effects as the fiber length at high power becomes comparable to the characteristic nonlinear length \cite{ zenteno1993high, kablukov2012output, schreiber2014analysis}. 

Different strategies have been investigated to improve the pump absorption in such cladding-pumping approach. For example, a Long-Period Grating (LPG) was used to couple the pump power injected in the cladding of the active fiber into the core of a 4-meter Yb-doped fiber cavity which led to an enhancement of pump absorption from 55\% to 80\%. The highest power achieved by this laser consequently increased by 55\%, from 4.67 to 7.27 W for 20 W of injected power. The LPG was UV-written in a photosensitive hydrogen-loaded single-mode and double-clad fiber using the point-by-point technique \cite{baek2006experimental}. Other approaches aimed to use all-fiber double pass pumping. A pump reflector was created by giving a right-angled conical shape to a passive single-mode fiber end with an electrical arc. The total internal reflection occurring at the fiber end led to a reflectivity of 55\% of the residual pump power. Adding such component increased the lasing efficiency with respect to the launched pump power from 30\% to 38\% resulting in an increase of the maximum laser output power from 2.1 to 2.7 W \cite{jeong2006simple}. This has, however, the drawback of preventing the use of fusion splices with protective endcap at the fiber end thus limiting the power scalability of such approach. In another attempt to increase the pump absorption in cladding-pumped lasers, a pump reflector made of a volume Bragg grating (VBG) was UV-written using the phase-mask inscription technique inside the hydrogen-loaded germanosilicate inner cladding of a triple-cladding speciality Yb-doped active fiber \cite{baek2004cladding}. A reflectivity of 46\% for the residual pump power was achieved at the pump wavelength of 916 nm by writing the VBG over the 42 \textmu m-diameter first inner cladding of such specialty fiber. The laser slope efficiency was increased from 21\% to 30\% with respect to injected pump power, whereas the maximum laser power went from 257 to 370 mW. UV-inscribed Bragg components have, however, the downside of requiring materials with enhanced photosensitivity, which greatly limits the gain fiber design. Femtosecond writing of Bragg structures would therefore be of great interest as it was shown to write highly stable and strong fiber Bragg gratings (FBG) directly in pure silica and other non-photosensitive materials \cite{mihailov2004bragg, bernier2007bragg, bernier2012writing, mihailov2011bragg}. This writing approach was also used to write VBGs in bulks of pure silica to stabilize the emission wavelength of 969 nm diode bars at the 4th order of the grating with no significant spectral drift being observed \cite{ voigtlander2011inscription,richter2017minimizing}. Femtosecond writing with a 400 nm beam was also demonstrated which opened the possibility of using such technique to inscribe fundamental FBGs operating in the short range of near-IR \cite{bernier2009ytterbium}. This portion of the spectrum is especially interesting as it encompasses the wavelengths of most diodes used to pump rare-earth doped silicate fiber lasers. 

In this paper, we report what we believe is the first demonstration of an all-fiber cladding-pumped laser schematic that integrates a femtosecond written VBG in the pure silica inner cladding of the active fiber to recycle the residual pump power. The usefulness of such component is demonstrated by optimizing the performances of a 1.6 \textmu m  erbium-doped silicate fiber laser operating at 25 W output power level. Femtosecond pulses at 400 nm are used along with the phase-mask writing technique to directly inscribe a chirped VBG-based pump reflector at the end of the gain fiber that yields 53\% peak reflectivity at the 976 nm pump wavelength guided in the highly multimode fiber’s cladding. Adding such component has the effect of increasing the lasing efficiency from 17.1\% up to 22.7\%, thus increasing the maximum output power from 20.8 W to 25 W, with the additional benefit of reducing by 25\% the optimal gain fiber length. 

The VBG is inscribed by using the same setup as the one described in \cite{habel2017femtosecond} based upon the phase-mask writing technique and femtosecond laser exposure \cite{mihailov2011bragg,thomas2007inscription, bernier2009ytterbium}. The writing beam is generated by a Ti-Sapphire regenerative amplifier system (Astrella, Coherent Inc.). It emits 806 nm pulses at a repetition rate of 1 kHz with a maximum output energy of 6 mJ. The transform-limited pulses duration is $\sim 30$ fs as measured by using a single-shot autocorrelator. The beam is then sent through a BBO crystal (EKSMA optics, BBO 1502) to generate a second-harmonic signal at 403 nm made of $\sim 40$ fs pulses with an energy that can reach 3 mJ. Both laser signals are separated by using a dichroic mirror in order to send only the 403 nm beam towards the inscription setup. The laser beam is focused by an acylindrical lens with a short focal length of 10 mm optimized for focusing with reduced aberration at 400 nm. Such lens is fixed on a 2D-piezoelectric stage that moves the laser beam across a maximum range of $\sim 100$  \textmu m x 100 \textmu m which corresponds to a conical-shaped writing area of $\sim 50$  \textmu m x 50 \textmu m in the fiber. The writing beam is also scanned along the fiber axis with a translation stage that moves the lens with regards to the fiber and the phase-mask which are kept fixed in close proximity of each other. The interference pattern is created by in-house fabricated chirped phase-masks with a central pitch of 674 nm and various linear chirp rates. The decoated portion of the fiber where the pump reflector is written is annealed at 475 $^{\circ} C$ for 10 minutes  after inscription to eliminate the photoinduced absorption losses in the doped fiber core resulting from the intense laser exposure \cite{ bernier2009ytterbium}.

\begin{figure}[b]
\centering
\includegraphics[width=1\linewidth]{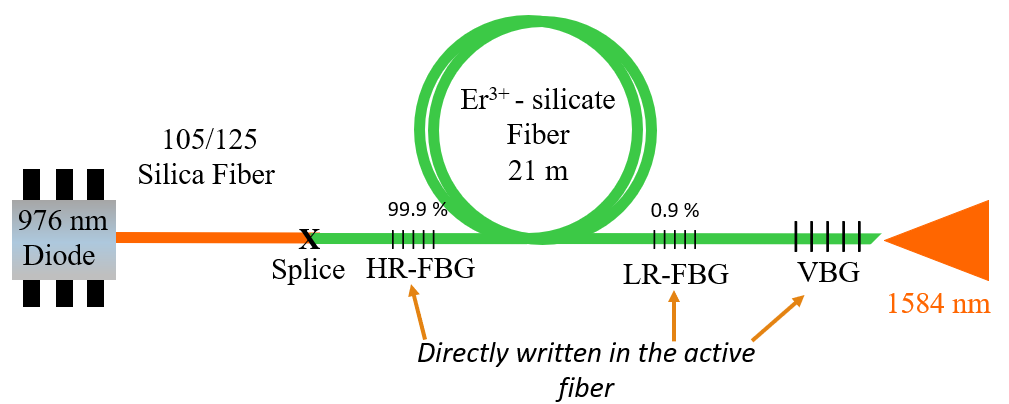}
\caption{Schematic of the Er-doped silicate all-fiber laser operating at 1.584 \textmu m. HR, high reflector; LR, low reflector; FBG, fiber Bragg grating; VBG, volume Bragg grating.}
\label{fig:cavite}
\end{figure}

The performances of the pump reflector are studied by writing the VBG directly in the active fiber of an all-fiber laser. This laser is made out of an in-house manufactured double-D shaped Er-doped silicate fiber having a core diameter of 17 \textmu m and a NA of 0.061 with a pure-silica inner cladding with diameters of 120 \textmu m x 130 \textmu m. The active fiber core is co-doped with erbium, aluminum and phosphorus to help reducing the occurrence of clustering while maintaining a small numerical aperture to ensure single-mode operation \cite{jebali2014264}. The laser setup seen in figure \ref{fig:cavite} is similar to the one used by \cite{pleau201820}. 

The laser is cladding-pumped by a fiber-coupled and wavelength stabilized 976 nm diode (nLight, Element, e18) providing a maximum CW pump of 120 W. The diode pigtail is spliced to a 21 m laser cavity bounded by two FBGs directly written in the Er-doped-fiber and inscribed through the coating \cite{habel2017femtosecond} with the 800 nm beam described previously. The high reflectivity input coupler (HR-FBG) has a peak reflectivity of 99.9\% over a broad bandwidth of 3 nm while the output coupler (LR-FBG) has a 0.9\% narrowband reflectivity to maximize the cavity performances. The VBG is inscribed directly in the inner cladding of a decoated segment of the active fiber after the output coupler, near the laser output. After inscription and annealing, low-index fluoroacrylate polymer is applied around the pump reflector that is UV-cured in a V-grooved copper block to ensure an efficient conductive thermal dissipation. It should be noted that no active cooling systems are used in the experiment. Finally, the fiber end at the laser output is angled cleaved in order to prevent parasitic lasing. 

\begin{figure}[b]
\centering
\includegraphics[width=1\linewidth]{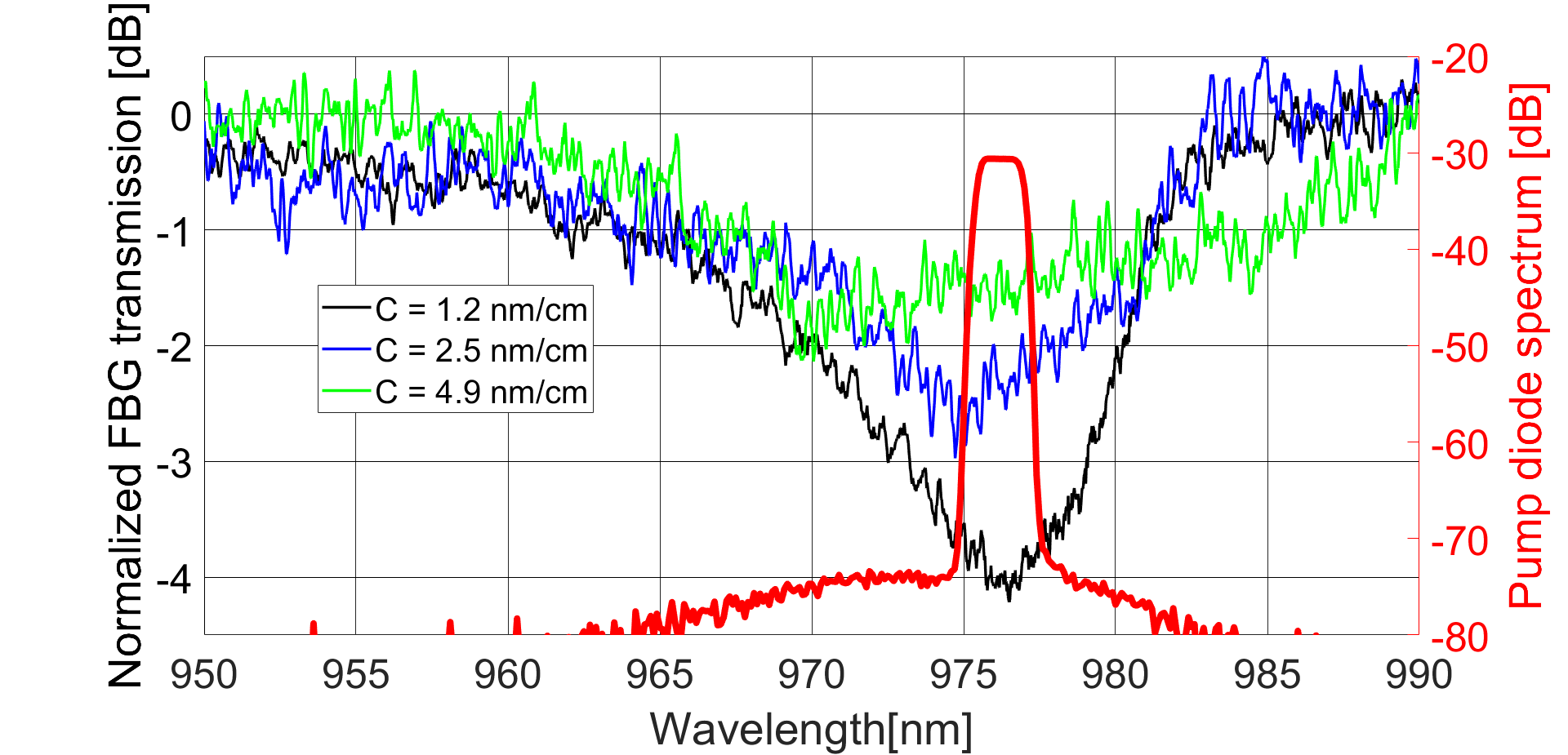}
\caption{Pump diode spectrum at high power long with the transmission spectra of three different pump reflectors written with the same inscription parameters but with different phase-masks with chirp rates of 1.2, 2.5 and 4.9 nm/cm.}
\label{fig:chirp}
\end{figure}

Before writing the final VBG in the cavity, optimization of the chirp rate for the phase-mask had to be conducted. The central wavelength and bandwidth of the pump reflector had to match those of the pump diode used. Since the pump propagation in the inner cladding of the Er-doped fiber is highly multimode (V $\sim 185$), the VBG has to be chirped enough such that it can interact with most of the pump modes having different refractive indices. Figure \ref{fig:chirp} shows the emission spectrum of the pump diode as well as the transmission spectrum of 3 VBGs written with the same nominal writing parameters but with different chirp rates. They were all inscribed with 140 \textmu J pulses for a 25 minute exposure time with a translation length of 26 mm along the mask and the fiber. They were then thermally annealed at 475 $^{\circ}$C for 10 minutes. As expected, the width of the transmission peak is increased as the chirp rate increases and the peak reflectivity decreases accordingly. The phase-mask with a chirp rate of 1.2 nm/cm was chosen as it resulted in a VBG that matched well the pump spectrum and offered the highest reflectivity. Further optimization could be done with reduced chirp rates and longer VBGs in order to reach a stronger peak reflectivity.  

\begin{figure}[t]
\centering
\includegraphics[width=1\linewidth]{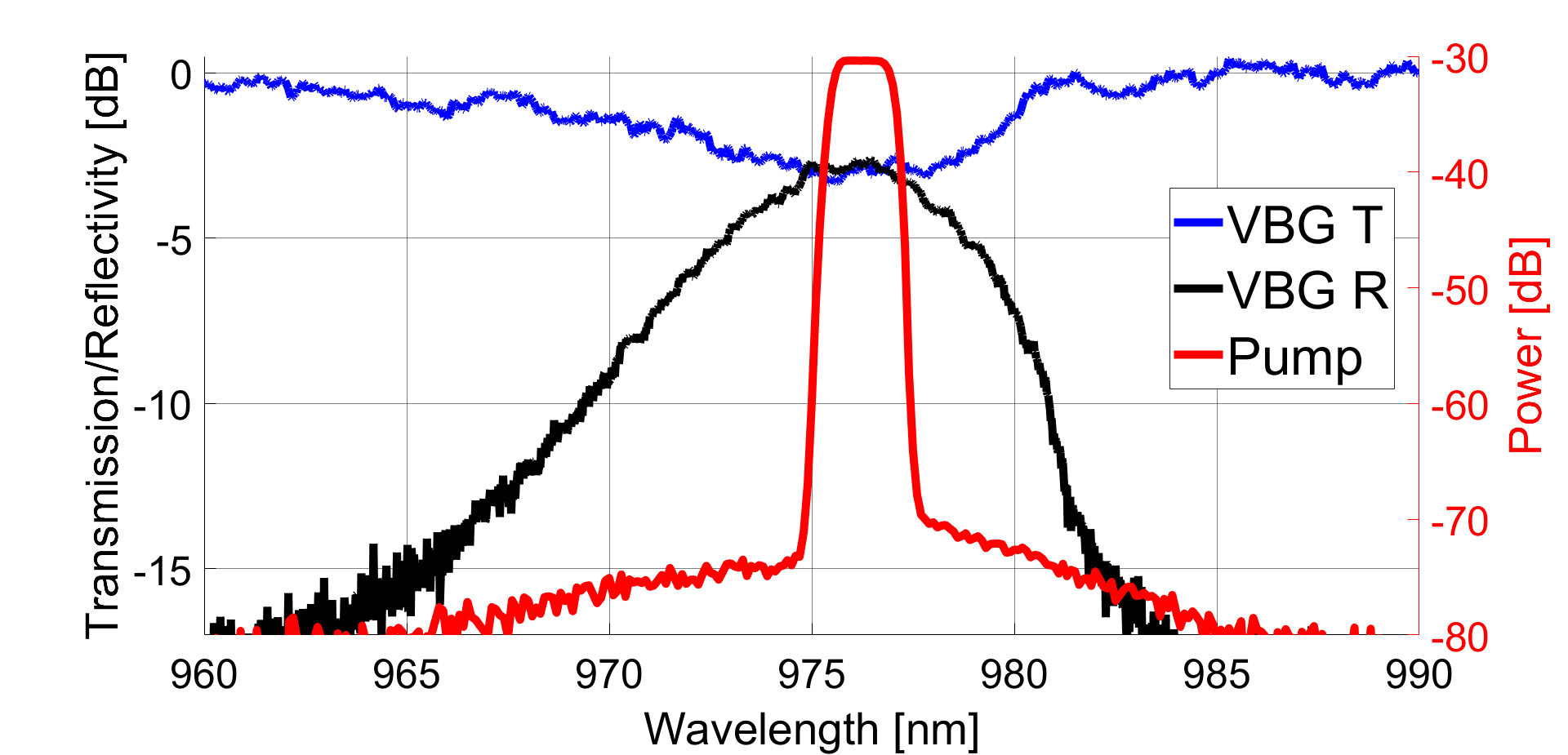}
\caption{Pump diode spectrum at high power along with the transmission and reflectivity spectra of the VBG written in the Er-doped final fiber laser cavity.}
\label{fig:vbg_final}
\end{figure}

Once the phase-mask chirp rate fixed at 1.2 nm/cm, the other writing parameters such as the pulse energy, the exposure time and the grating length had to be optimized to reach the strongest pump reflectivity while not inducing significant core signal losses. The transmission spectrum of the final VBG used in the laser cavity before being recoated is shown on figure \ref{fig:vbg_final}. It was written with 200 \textmu J pulses for 30 minutes and with a translation length of 26 mm. It was also thermally annealed at 475 $^{\circ}C$ for 10 minutes. An insertion loss of -3.2 dB is measured at 976 nm from which we can infer a peak reflectivity of about 53 \% without measurable pump losses evaluated at wavelengths above the Bragg resonance. Also, no significant signal losses ($\leqslant 0.1$ dB limited by the measurement setup) near 1.6 \textmu m could be measured by cutback.

\begin{figure}[b]
\centering
\includegraphics[width=0.5\linewidth]{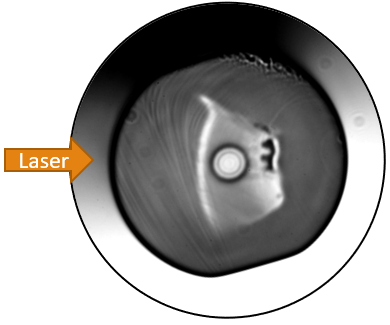}
\caption{Phase-contrast microscope image of the cross section of a VBG written with the same inscription parameters as the one inscribed in the final cavity. The orientation of the fiber with respect to the laser beam during the inscription is given by the direction of the arrow.}
\label{fig:micro}
\end{figure}

The influence of the fiber’s curvature on the shape of the writing surface can be seen on figure \ref{fig:micro} where a fiber with a VBG written under similar conditions as the VBG presented in figure \ref{fig:vbg_final} was cleaved and analyzed with a phase-contrast microscope. As the beam is scanned during the inscription, it is refracted at different angles depending on its incident height on the fiber. This usually does not have a significant impact on the writing of FBGs since the core size is much smaller than the cladding curvature. For VBGs written in the cladding of the fiber, it has, however, the effect of reducing the writing surface and distorting its shape into a conical profile. 

Hot spots where more energy was deposited during the writing can also be seen in figure \ref{fig:micro} as black regions on the right side of the grating. It was observed that those black spots were not inducing pump losses when they were written in the inner cladding as in that case. However, they could lead to important signal losses if they were overlapping the doped core region. Those hot spots are obtained if the beam is not exclusively incident on the curved part of the cladding during the scan. In fact, if the beam is scanned at the edge of the curved and straight surfaces of the fiber, hot spots will appear at the intersection of those two. The fiber orientation and alignment are therefore critical when a non-circular fiber is used. For the final VBG shown in figure \ref{fig:vbg_final}, particular care was taken to properly align the fiber to prevent this effect from happening. The writing surface is highly dependent on whether the beam is incident on a straight surface or a curved surface.

\begin{figure}[t]
\centering
\includegraphics[width=1\linewidth]{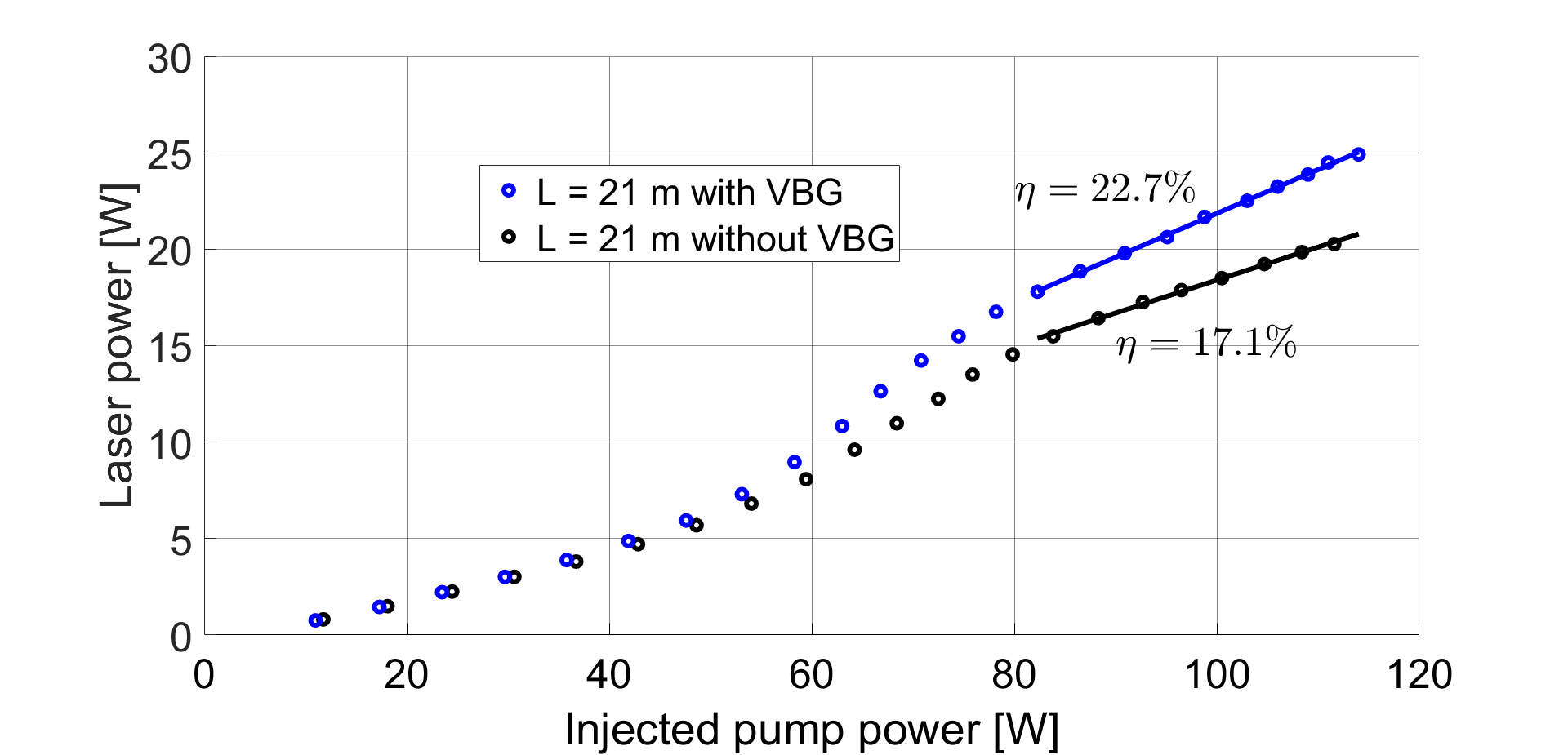}
\caption{Laser output power of the Er-doped fiber laser as function of the injected pump power with and without the presence of a VBG. The pump reflector increases the slope efficiency from 17.1\% to 22.7\% at high pump power when the diode wavelength is stabilized at 976 nm.}
\label{fig:eff}
\end{figure}

The performances of the laser cavity with and without the final VBG are shown in figure \ref{fig:eff}. The two curves have a nonlinear shape due to the fact that the emission wavelength of the pump diode shifts as the power is increased. It changes from 962 nm at low power and gradually locks to 976 nm at output power around $\sim 80$ W. The laser efficiencies from those curves are therefore only meaningful at high power when the pump spectrum matches the reflectivity spectrum of the VBG. One can see that the presence of the pump reflector increased significantly the cavity efficiency with respect to the launched pump power from 17.1\% up to 22.7\%, as well as the maximum output power from 20.8 W to 25 W. 

\begin{figure}[t]
\centering
\includegraphics[width=1\linewidth]{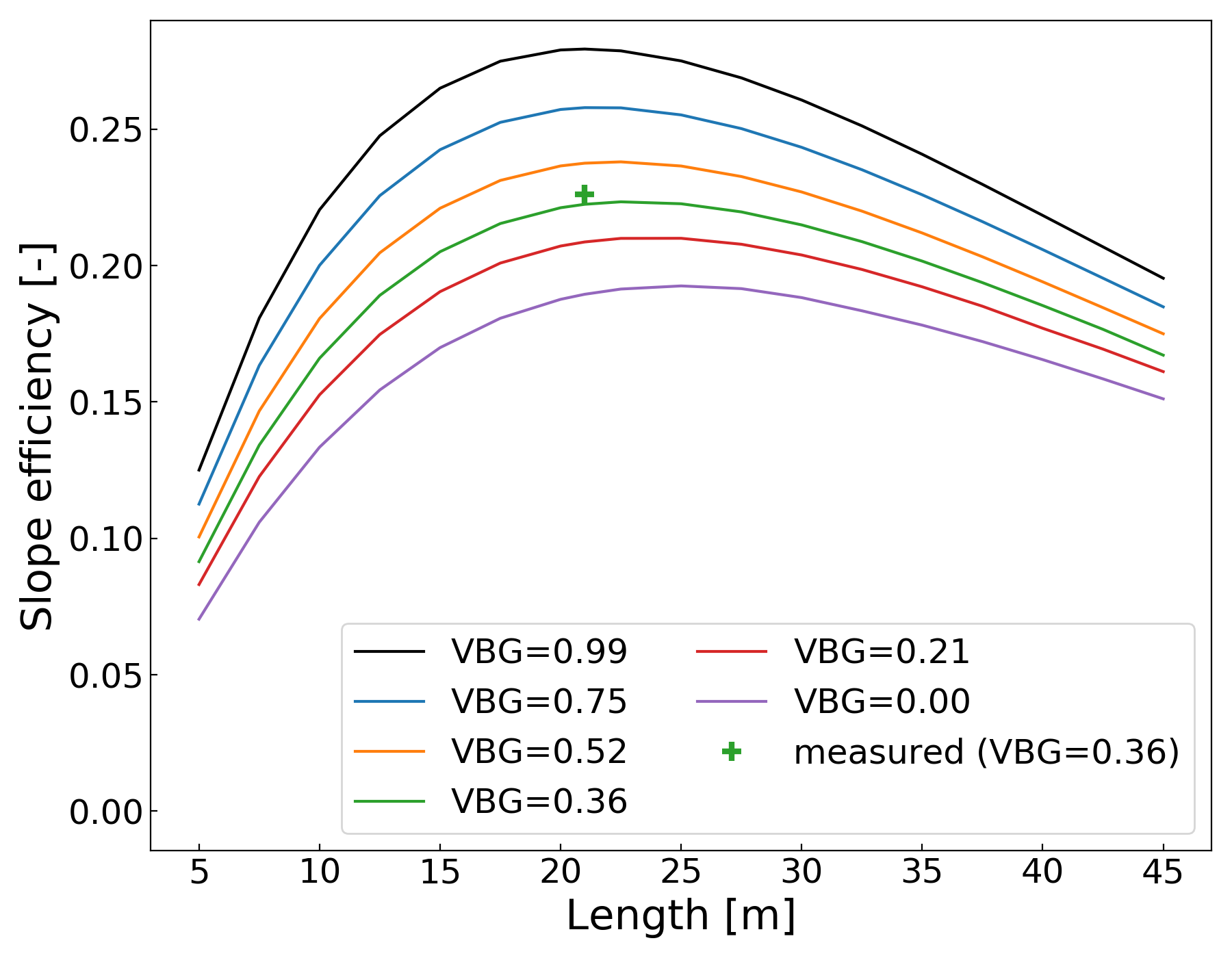}
\caption{Simulations of the influence of the VBG reflectivity on the slope efficiency with regards to the cavity length.}
\label{fig:sim}
\end{figure}

Numerical simulations of the laser cavity were performed using a similar model to the one presented in \cite{pleau201820} to evaluate the effective VBG reflectivity at high pump power and to study the influence of the reflectivity value on the laser performances. Figure \ref{fig:sim} shows the evolution of the expected laser slope efficiency for various VBG reflectivities as function of the cavity length. The green cross indicates the measured performances and indicates that the effective reflectivity would be about 36\%, a value significantly lower than the 53\% retrieved from the transmission measurement. This points out that the VBG does induce slight pump and/or signal losses that could be either absorbed and/or outcoupled from the fiber numerical aperture to radiation modes. The simulations also show that with a 21 m meter cavity and a VBG with an effective reflectivity of 99\%, the lasing efficiency could be increased up to 28\% which confirms the great benefits of such component for the laser performances. These calculations also indicate that the optimal cavity length without VBG is 28 m. Adding an optimized VBG as pump reflector reduces this optimal length to 21 m which allows for a direct reduction of the cavity length (cost) by 25\%.

\begin{figure}[b]
\centering
\includegraphics[width=1\linewidth]{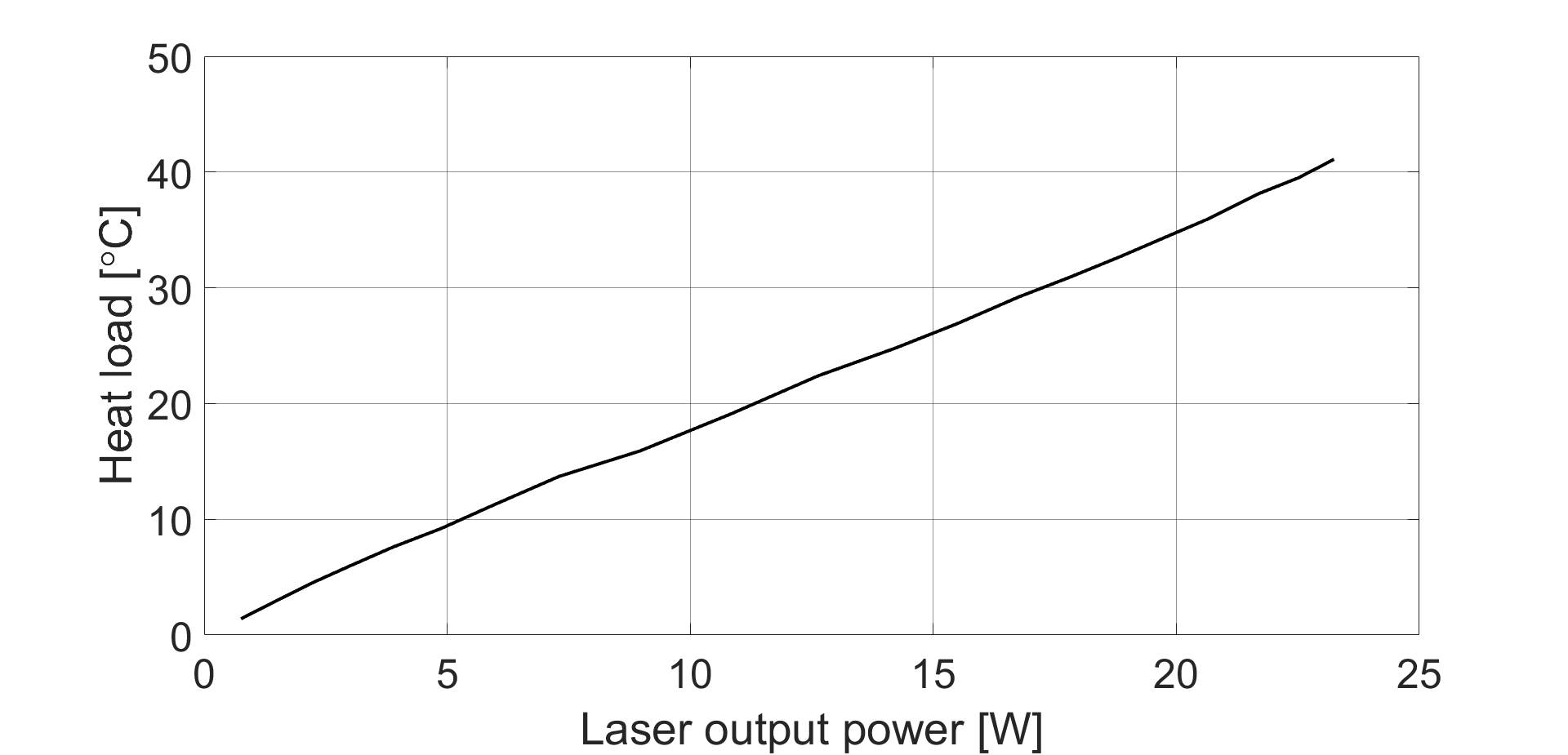}
\caption{Heat load on the VBG as function of the laser power.}
\label{fig:temp}
\end{figure}

The temperature of the VBG and the copper block on which it was fixed by a recoating polymer was then measured with a Jenoptik IR camera. According to figure \ref{fig:temp}, the laser signal of 23 W induced a heat load of 44 $^{\circ} $C on the pump reflector. By splicing the VBG directly to the delivery fiber of the pump, it was found that the heat load was significantly less than with the laser power. This indicates that the pump reflector has induced core signal losses which is an issue that will need to be addressed before using such a component for higher power systems.

The transverse dimensions of the VBG, and thus its maximum reflectivity, could be increased by using piezoelectric actuators with an enhanced range for the 2D-scanning of the acylindrical lens. The influence of the fiber’s curvature on the refraction of the writing beam could then be limited by inserting the fiber into a hollow capillary with a much larger outside diameter or by using an active fiber with a larger inner-cladding diameter. With the increased reflectivity of the VBG, less energetic pulses could be used which would reduce the induced losses in the core. Another strategy to mitigate these limitations would be to write such pump reflector in the passive fiber generally used as a delivery fiber which is undoped and of circular geometry.

In summary, we propose a new all-fiber cladding-pumped laser schematic that makes use of a femtosecond written chirped VBG as a pump reflector in the pure-silica inner cladding of the gain fiber. This allows reinjecting part of the residual pump power after a single pass back into the cavity which increases the pump absorption and redistributes more homogeneously the gain along the cavity. Such a component was written directly in the active fiber of an Er-doped fiber laser. According to its transmission spectrum, the chirped VBG written in the Er-doped fiber has a peak reflectivity of 53\% which managed to increase the laser efficiency by 33\% from 17.1\% up to 22.7\%. The maximum output power was increased from 20.8 W to 25 W while the optimal cavity length was reduced by 25\%. Those results demonstrate the relevance of such all-fiber pump reflector to reduce significantly high-power cladding-pumped laser cavity lengths with optimized performances. \\

\noindent Canada Foundation for Innovation (CFI); Fonds de recherche du Québec—Nature et technologies (FRQ.NT) (FT114976); Natural Sciences and Engineering Research Council of Canada (NSERC) (CG112389).\\

\noindent The authors would like to thank Marc D’Auteuil and Stephan Gagnon for technical assistance, as well as Tommy Boilard, Marie-Pier Lord and Frédéric Maes for helpful discussions.

\noindent










\bibliography{sample}

\bibliographyfullrefs{sample}


\ifthenelse{\equal{\journalref}{aop}}{%
\section*{Author Biographies}
\begingroup
\setlength\intextsep{0pt}
\begin{minipage}[t][6.3cm][t]{1.0\textwidth} 
  \begin{wrapfigure}{L}{0.25\textwidth}
    \includegraphics[width=0.25\textwidth]{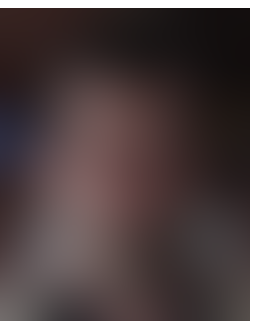}
  \end{wrapfigure}
  \noindent
  {\bfseries John Smith} received his BSc (Mathematics) in 2000 from The University of Maryland. His research interests include lasers and optics.
\end{minipage}
\begin{minipage}{1.0\textwidth}
  \begin{wrapfigure}{L}{0.25\textwidth}
    \includegraphics[width=0.25\textwidth]{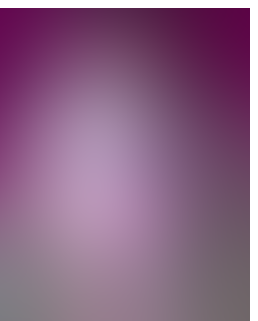}
  \end{wrapfigure}
  \noindent
  {\bfseries Alice Smith} also received her BSc (Mathematics) in 2000 from The University of Maryland. Her research interests also include lasers and optics.
\end{minipage}
\endgroup
}{}

\end{document}